\documentclass{article}
\pdfoutput=1 

\usepackage[margin=1.0in]{geometry}
\usepackage{lineno}

\usepackage{amssymb}
\usepackage{amsthm}

\usepackage{mathtools}
\usepackage{amsfonts}

\usepackage{subcaption}

\usepackage{hyperref}

\newcommand{\be}{\begin{equation}}
\newcommand{\ee}{\end{equation}}

\title{Structure of vertices in massless theories}

\author{ S. Srednyak ,Duke University, Durham, USA}

\begin{document}
\maketitle

\begin{abstract}
We characterize the singularity set of massless theories by giving a complete set of the Landau polynomials. We find the general form of Gauss-Manin connection. We show that for massless theories the dependence on momenta decouples from the dependence on coupling and dimension. The latter is completely absorbed into a set of matrices that have no dependence on kinematic variables.
\end{abstract}

\tableofcontents

\section{Introduction}

There has been considerable interest in deriving differential equations for Feynman integrals. The famous epsilon-form \cite{Henn:2013pwa} was used in papers \cite{adams2014two} to discuss relationships with modular forms ( see also \cite{Bloch:2016izu}). At two loops there is a large body of literature including \cite{ Canko:2020ylt, Chen:2020iqi, Frellesvig:2019byn, Bonciani:2019jyb, Heller:2019gkq, Becchetti:2019tjy, Chaubey:2019lum, Becchetti:2017abb,Bonciani:2016qxi, Bonciani:2016ypc, Ita:2015tya, Remiddi:2013joa, Gehrmann:2014bfa}. There is relatively less literature on master integrals in pure glue theory, although more complicated massive intergals were considered e.g. in  \cite{DiVita:2017xlr, Bonciani:2016ypc}. The computation of cusp anomalous dimensions was a major focus \cite{vonManteuffel:2020vjv,   Huber:2019fxe, Henn:2019swt, Bruser:2019auj, Moch:2018wjh, Grozin:2017css, Henn:2016men}. Master integrals always exist \cite{griffiths1969periods} and have remarkable block structure as dictated by toric geometry of the integrand. In physics they were usually derived by Laporta algorithm \cite{laporta2000high}. This algorithm , together with its algebro-geometric interpretation, was known long before in the mathematical literature ( see refs in \cite{gelfand1994discriminants}, where it is traced at least to Cayley).  Recently there have been interesting developments in conversion of the GZK D-module to the GM connection \cite{hibi2017pfaffian}. The subject of computation of Feynman integrals may be considered as realising a particular case of hyperfunction analysis \cite{beilinson2018faisceaux}. In the context of 2-loop integrals, there emerged large literature on master inntegrals, as this is the next theoretical frontier for LHC phenomenology. Some of the papers that we found useful include \cite{Badger:2022hno,Abreu:2022vei,Duhr:2021fhk,Canko:2020ylt,Chaubey:2019lum,Frellesvig:2019byn,Heller:2019gkq,Becchetti:2019tjy,Bonciani:2016qxi,Ita:2015tya}. In these papers the computation of master integrals relies on numerical algorithms.

In this paper , we find that in the case of massless theories the general equations simplify substantially. At the root of this simplification lies the fact that there exists a convenient explicit form for the  Landau polynomials. These polynomials are given by Gram determinants of sums of external momenta. We then use the flatness of the Gauss-Manin connection to infer our differential equations both on diagram level and for the sum over diagrams. Our result allows reduction of the computation of Feynman diagrams to the combinatorial problem of determining certain matrices as series of the coupling and dimension and the solution of the system of differential equations.

Section \ref{sec:results} states our main results. The proofs are given in Section \ref{sec:proofs}.

\section{Aknowledgemets}

The author was supported in part by  BNL LDRD 21-045S NPP.

\section{Preliminaries}

In the following we are concerned with the usual formulation of perturbation theory for massless gauge theory \cite{Bogner:2007mn}. We are focusing on the integrals
\be
J_m = \int (dq) \frac{q^m}{\prod D_i(q,p)}
\ee
where $q =(q_{a,\mu})$ denotes a point in loop momentum space. We assume that Wick rotation was already performed. The integration happens over the class $[\mathbb{RP}^{dL}] \in H_{dL}(\mathbb{CP}^{dL} -\cup \{D_i =0 \};\mathbb{CP}^{dL-1} - \cup \{D_i =0 \} \cap \mathbb{CP}^{dL-1})$.

Throughout the paper we use dimensional regularization which corresponds to replacement
\be
(dq) = \wedge dq_\alpha  \rightarrow \frac{1}{x_0^{dL+1}} \sum_{\alpha} (-1)^\alpha dx_0\wedge ... \hat{dx_\alpha} ...\wedge dx_{dL}
\ee
where $\alpha= (a,\mu)$ and $x_\alpha$ are the homogeneous coordinates on the projectivised loop momentum space.

We also use the following notation 
\be
\Delta(p_1,p_2,...) = Gram(p_ip_j)
\ee
for the Gram determinant of the d-vectors $p_i$.

\section{Results}
\label{sec:results}

In this section we formulate our results.

{\it 
{\bf Th0:} (Singularities of the triple vertex) The set of Landau polynomials of the triple vertex is $p^2,p'^2,\Delta(p,p')$.
\qedsymbol{}

{\bf Th0.1:} (Singularities of the quartic vertex) The set of Landau polynomials of the quartic vertex is $\Delta(p_i),\Delta(p_i,p_j),\Delta(p_1,p_2,p_3)$.
\qedsymbol{}

{\bf Th0.2:} Landau varieties of any diagram function in a massless theory are among $\Delta(p'_i)$ , $\Delta(p'_i,p'_j)$ , $\Delta(p'_i,p'_j,p'_k)$ , $\Delta(p'_i,p'_j,p'_k,p'_l)$ and $p'_i = \sum \pm p_k$, where $p_k$ are external momenta.
\qedsymbol{}

}

{\it
{\bf Th1:} ( Structure of triple vertex). For any diagram D contribution $V_D(p,p')$ to the triple vertex of pure glue QCD the following holds true
\be
\frac{\partial U_D}{\partial z_i} = ( \frac{A^D_i}{p^2} + \frac{B^D_i}{(p')^2}+ \frac{C^D_i}{(p+p')^2}  + \frac{D^D_ip^2+E^D_i(p')^2+F^D_ipp'+G^D_i}{\Delta(p,p')} )U_D
\ee
where $z_i = p^2,p'^2,pp'$. Matrices $A^D_i,B^D_i,C^D_i,D^D_i,E^D_i,F^D_i,G^D_i$ are functions of dimension d and the non renormalized coupling and the diagram topology only.

If we sum over the diagrams , then there exists (infinite dimensional) vector function $U(p,p')$ and matrices $A_i$ such that
\be
\frac{\partial U}{\partial z_i} = ( \frac{A_i}{p^2} + \frac{B_i}{p'^2}+ \frac{C_i}{(p+p')^2} + \frac{D_ip^2+E_ip'^2+F_ipp'+G_i}{\Delta(p,p')})U
\ee
with the notations above, and where the vertex is included as the highest weight component of this vector.
\qedsymbol{}

}

{\bf Property (*)} We say that a quantity $w$ has property (*) if it is a function of dimension and unrenormalized coupling and diagram topology only.

{\it
{\bf Th2:} (Structure of 4-vertex). For any diagram D contributing to the 4-vertex $X^D(p_1,p_2,p_3)$ of pure glue QCD the following holds true
\be
\frac{\partial X^D}{\partial z_k} = ( \sum_i \frac{A^D_{k,i}}{p_i^2} + \sum_{i,j} \frac{B^D_{k,i,j}}{\Delta(p_i,p_j)} +
 \frac{C^D_{k}}{\Delta(p_1,p_2,p_3)}) X^D
\ee
where $A^D_{k,i}$ have property (*), $B_{k,i,j}$ are linear functions of $p_ip_j$ with coefficients having property (*), and $C_{k}$ are quadratic functions of $p_ip_j$ with coefficients having property (*). Similar formula holds for the sum over all diagrams. 
\qedsymbol{}
}

The structure of the flat connection for arbitrary diagram is given as

{\it 
{\bf Th3:} For arbitrary diagram, the diagram function is included into a flat connection J such that the following holds
\be
\frac{\partial J}{\partial z_a} = (\sum \frac{A_{I,a}}{L_I} ) J
\ee
where $L_I$ are the Landau polynomials of the diagram, $z_a=p_ip_j$ or $z_a=m_i^2$ are the kinematic variables, and $A_I$ are certain matrices that depend polynomially on kinematic variables, coupling and dimension.
\qedsymbol{}
}

\section{Proofs}
\label{sec:proofs}

\subsection{Proof of Th0}

Singularities of a diagram function occur when there are vanishing cycles in the complement of propagators. This  means the following. There exists a group of propagators $D_{i_s},s=0..K$ such that there is a vanishing cycle in the intersection $\cap D_{i_s}$ but not in any intersection of any subgroup of these propagators.  For definiteness, we will focus on the main singularities where the above condition determines all the loop momenta. We will see shortly that this implies $K=2L$. The condition for the existence of the vanishing cycles is the existence of a non trivial linear relationship between the normals
\be
\sum v_s \frac{\partial D_{i_s}}{\partial q_{a,\mu} } =0
\ee
where not all $v_s$ are 0. We have to be careful in this statement because the quadrics $D_i$ are highly singular. In particular, they have $(d-1)L$-dimensional locus where their derivative is identically zero. Furthermore, there is an additional point - the conical point - where there is additional singularity ( the quadrics are in fact cones over $d-1$-dimensional nonsingular quadrics). Nonetheless, the above condition still holds, even in the case where vanishing happens at some of the conic points of some subset of the propagators. This is a consequence of general theory of vanishing cycles for possibly singular algebraic varieties and schemes \cite{kashiwara2013sheaves, hotta2007d,gelfand1990generalized }

The above condition can be solved ( for main singularities) in the form
\be
q_a = \alpha_a p+\beta_a p'
\ee
where $\alpha_a,\beta_a$ are rational functions of $v_s$. It can be seen that instead of $v_s$, for main singularities we can take $\alpha_s,\beta_s$ as parameters instead of $v_s$, i.e., there is generically 1-1 map between spaces of $v_s$ and $\alpha_s,\beta_s$. We will do this in the following.

For determination of Landau polynomials, it is enough to consider the generic point of the $p,p'$ space. Therefore, we may assume that we use the $SL(4,C)$ symmetry ( still in signature (+,-,-,-), for simplicity) of the system to choose coordinates in which
\begin{gather}
p=(r,0,0,0) \\
p'=(s,t,0,0)
\end{gather}
The impossibility of such choice is equivalent to $p^2=0$ , or $p'^2=0$, which are clearly among the Landau varieties. We will assume that such a choice has been made. From this equation ,we see that
\be
q_i = (q_{i,0},q_{i,1},0,0)
\ee

Then the condition $D_i=0$ splits into
\be
q_{i,0} = \pm q_{i,1}
\ee
We now use the fact that
\be
q_{i,0} = (\sum l_{i,a} \alpha_a )r + (\sum l_{i,a} \beta_a ) s
\ee
and
\be
q_{i,1} =(\sum l_{i,a} \beta_a )t
\ee
This homogeneous in $\alpha_a,\beta_a$ system has nonzero in $\alpha_a,\beta_a$ solution only if $t=0$, which is equivalent to $\Delta(p,p')=0$, which proves the theorem.

\subsection{Proof of Th0.1 and Th0.2.}

The above proof does not generalize to the 4-vertex as there are 3 independent vectors of momenta. We therefore develop an alternative inductive proof. We will prove a little more general theorem Th0.2. It is based on the following basic statement

{\it 
{\bf Lemma:} The following equality holds
\be
\Delta(p_1,...,p_n,q) = \Delta(p_1,...,p_n)q^2_\perp
\ee
where $p_i$ are arbitrary d-dimensional vectors in $\mathbb{C}^d$ and $q_\perp$ is the component of $q$ orthogonal to the space spanned by $p_i$ in the metric $p^2=p_1^2+...+p_d^2$.
}

This lemma has the following geometric significance

{\it 
{\bf Lemma:} The variety
\be
\{q: \Delta(p_1,...,p_n,q)=0\}
\ee
is a product of n-dimensional linear space spanned by $p_i$ and the (d-n-1)-dimensional conic  $\{q: q_\perp^2 =0\}$ where $q_\perp$ is the orthogonal complement to the above linear space in the metric $q^2=q_1^2+...+q_d^2$.
}

We will also need the following

{\it
{\bf Lemma:}
\be
\Delta(p_i+a_iq,...,p_j+a_jq,b+q) = \Delta(p_i-a_ib,...,p_j-a_jb,q)
\ee
for any scalalrs $a_k$ and d-vector b.
}

We will carry out the proof induction in the number of loops L. At L=0 the statement is true because the only singularities are $p'^2=0$ where $p'$ is any of the momenta that flow through internal lines of a tree diagram. Suppose the statement holds for L-1. To prove it for L, we need to analyse the vanishing cycles that occur in the integral
\be
f_L(p_{ext}) = \int dq P(p,q) f_{L-1}(p_{ext},q)
\ee
where $P(.)$ is the product of zero or more propagators. By $f_l$ we denoted an l-loop diagram function. By induction hypothesis, these vanishing cycles must occur in a set of varieties $\Delta(p'_i,...,p'_j)=0$ where $p'$ are the linear combinations of $p_{ext}$ and $q$. For definiteness, we consider vanishing cycles in the system of only two varieties ( we can call this situation 2-pinch). For example we restrict the varieties $\Delta(p'_i,p'_j,p'_k)=0$ and $\Delta(p'_l,p'_m)=0$. Due to the above lemmas, the equations for them can be transformed to the case
\begin{gather}
\Delta(r_1,r_2,q)=0 \\
\Delta(r_3,q)=0
\end{gather}
where $r_i$ are linear combinations of external momenta with coefficients $\pm 1$. These varieties have the form
\begin{gather}
L(r_1,r_2)\times D_1 \\
L(r_3)\times D_2
\end{gather}
where $D_1$ is a conic in the complement to the span of $r_1,r_2$ and $D_2$ is the conic in the complement to the span of $r_3$. Vanishing cycles in the complement to these to varieties happen only if $\Delta(r_1,r_2,r_3)=0$ ( assuming $\Delta(r_i)\neq 0, \Delta(r_i,r_j) \neq 0$ ), which proves the inductive hypothesis in this case.

The general case is analysed analogously, with the difference that there are several conics in the complement to linear spaces spanned by $p'_i$ with appropriate linear combinations of external momenta.

\subsection{Proofs of Th1,Th2,Th3}

In this section we show how Th1,Th2 follow from Th0-Th0.2. The proof is based on the following

{\it 
{\bf Lemma:} Consider flat connection for function $f(x_1,...,x_n)$
\be
d_if = \Omega_i f
\ee
where $\Omega_i$ is a rational function of $x_j$. Suppose that $\Omega$ has singularities only at zeros of $P_I$ where $P_I$ are polynomials. Suppose further that the singularities of $f$ are regular, i.e. near each component $l$ of singularity locus the function behaves as
\be
f =L^\alpha R_1+R_2
\ee
where $R_i$ are regular functions. Then
\be
\Omega_i = \sum \frac{A_{I,i}}{P_I}
\ee
where $A_{I,i}$ are some polynomials.
}

{\bf Proof}.

The fact that $\Omega$ are rational and singularities are among $P_I$ allows us to write
\be
\Omega_i =\sum \frac{A_{I_1,...,I_k,i} }{P_{I_1},...,P_{I_k} }
\ee
for polynomials $A_{I_1,...,I_k,i}$. All $I_s$ must be distinct by regularity. What remains to be proved is that they are zero except for the case $k=1$. We derive this from regularity conditions. Consider for example the terms
\be
\frac{A_i}{P_1P_2}
\ee
Introduce change of variables
\be
z_i = P_i, i=1,2
\ee
Then the first two components of the differential equations take the form
\be
\frac{\partial f}{\partial z_i} =( \frac{A_i}{z_1 z_2 } +...) f , i=1,2
\ee
where dots stand for terms that do not contain $z_1z_2$ in the denominator, although they may contain $z_1$ or $z_2$ only. Then from the above equation for $i=1$ we get
\be
f = (z_i)^{\lambda_1/z_2}R_1+R_2
\ee
where $\lambda_1$ is an eigenvalue of $A_1(0,0,....)$ and $R_i$ are regular in $z_1$. But such forms are not allowed by regularity. Therefore, we immediately get that $A_1(0,0,,...)$ is identically zero. This means that
\be
A_1 = P_1 B_1+P_2B_2
\ee
for some polynomials $B_i$.

What remains is to consider components of the differential equations for coordinates $x_2,...,x_n$.
\be
d_i f = (\frac{A_i}{z_1z_2}+...)f, i=2,...n,
\ee
where dots denote terms that do not contain products of terms $z_1z_2$ in the denominator. Near $z_1z_2=0$ we get
\be
f = C exp(\frac{\int A_i(z_1,z_2,x_2,...) dx_i }{z_1z_2} +...)
\ee
from which we infer that $A_i$ must be divisible by $z_1$ and $z_2$, by regularity. This proves the lemma.

\qedsymbol{}

{\bf Remark:}
The structure lemma proved above has wider applicability. Consider the integrals
\be
J = \int P^\alpha x^m d^dx
\ee
where $m_i$ are positive integers and $P(x)= \sum p_\omega x^\omega$ is a polynomial in the generic stratum corresponding to a Newton polytope. This is a classical problem considered e.g. in \cite{arnold2012singularities}. Then there are only two components of the singularity of $J$ - the discriminant $D$ of P and the discriminant $D_\infty$ of $P$ restricted to the compactification stratum $\mathbb{CP}^{d-1}$ of the x-space. Then from the above lemma we can infer that J is a part of flat connection $I$ that satisfies
\be
\frac{\partial I}{\partial p_\omega } = ( \frac{A_\omega}{D} + \frac{B_\omega}{D_\infty} ) I
\ee
for some polynomial matrices $A,B$.

This observation is already meaningful in the case $d=1$ when the integral is reducible to the Lauricella function
\be
J = \int ( p_n x^n +...+p_0)^a x
\ee
\be
J = (p_d)^a\int ( (x-x_1)...(x-x_n))^a x
\ee
In this case, the flat connection involves differences $x_i-x_j$ which after some calculation can be reduced to discriminant by introducing appropriate polynomial numerator.

{\bf Corr:} The theorems $Th1-3$ are corollaries of the above lemma and $Th0-0.2$.

\section{Discussion}

Our results allow reduction of the computation of Feynman diagrams to the combinatorial problem of calculating the matrices $A_i$ etc. that stand in the rhs of the GM connection. The dependence on momenta thus completely decoupled from the dependence on coupling and the complex dimension. We would like to point out several calculations that become immediately available.


The first concerns recursion for the matrices $A^D_i$ as functions of diagram. The vertex functions satisfy Schwinger-Dyson equations that lead to a recursive relation on $A_i$.


The simple form of the flat connection suggests that in the massless case there is significant simplification in interpreting the Bogoliubov-Shirkov second quantization equations. The Riemann-Hilbert approach provides an alternative and potentially simpler formulation in this case.


It is known that in some cases there are simple dimension shift formulas that coefficients $A_i(d,\alpha)$ satisfy with respect to the dimension. Our approach sheds new light on them in the massless case.


Th3 demonstrates that the complexity of Feynman diagram functions mainly lies in algebraic properties of the Landau polynomials.  For massive theories , the degree of these polynomials grows exponentially with the number of vertices  and there are no general algorithms to compute them.

\section{Conclusion}

In this paper we derived a flat connection for the vertices of massless theories. This connection has the characteristic feature that there are only a few rational terms with explicitly described polynomials in denominators. The computation of the connection therefore reduces to the computation of the numerators, which further reduces to computation of certain matrices that are independent of kinematic variables.

This result brings to the forth the question of algorithmic computation of the constants in the connection matrices. This problem is complicated because its solution must involve representation theoretic classification of bases of the flat connection, which in turn must be sensitive to the toric symmetry of the diagram.

Our representation has further implications on the Riemann-Hilbert formulation of the perturbation theory. In this case we have an RH problem with regular singularities. It completely specifies the singularity locus of the RH problem.

We hope that our representation will be useful for numerical evaluation of the diagrams.

\bibliographystyle{unsrt}
\bibliography{ref}

\begin{thebibliography}{10}

\bibitem{Henn:2013pwa}
Johannes~M. Henn.
\newblock {Multiloop integrals in dimensional regularization made simple}.
\newblock {\em Phys. Rev. Lett.}, 110:251601, 2013.

\bibitem{adams2014two}
Luise Adams, Christian Bogner, and Stefan Weinzierl.
\newblock The two-loop sunrise graph in two space-time dimensions with
  arbitrary masses in terms of elliptic dilogarithms.
\newblock {\em Journal of Mathematical Physics}, 55(10):102301, 2014.

\bibitem{Bloch:2016izu}
Spencer Bloch, Matt Kerr, and Pierre Vanhove.
\newblock {Local mirror symmetry and the sunset Feynman integral}.
\newblock {\em Adv. Theor. Math. Phys.}, 21:1373--1453, 2017.

\bibitem{Canko:2020ylt}
Dhimiter~D. Canko, Costas~G. Papadopoulos, and Nikolaos Syrrakos.
\newblock {Analytic representation of all planar two-loop five-point Master
  Integrals with one off-shell leg}.
\newblock {\em JHEP}, 01:199, 2021.

\bibitem{Chen:2020iqi}
Long-Bin Chen, Wei Wang, and Ruilin Zhu.
\newblock {Master integrals for two-loop QCD corrections to quark quasi PDFs}.
\newblock {\em JHEP}, 10:079, 2020.

\bibitem{Frellesvig:2019byn}
Hjalte Frellesvig, Martijn Hidding, Leila Maestri, Francesco Moriello, and
  Giulio Salvatori.
\newblock {The complete set of two-loop master integrals for Higgs + jet
  production in QCD}.
\newblock {\em JHEP}, 06:093, 2020.

\bibitem{Bonciani:2019jyb}
R.~Bonciani, V.~Del~Duca, H.~Frellesvig, J.~M. Henn, M.~Hidding, L.~Maestri,
  F.~Moriello, G.~Salvatori, and V.~A. Smirnov.
\newblock {Evaluating a family of two-loop non-planar master integrals for
  Higgs + jet production with full heavy-quark mass dependence}.
\newblock {\em JHEP}, 01:132, 2020.

\bibitem{Heller:2019gkq}
Matthias Heller, Andreas von Manteuffel, and Robert~M. Schabinger.
\newblock {Multiple polylogarithms with algebraic arguments and the two-loop
  EW-QCD Drell-Yan master integrals}.
\newblock {\em Phys. Rev. D}, 102(1):016025, 2020.

\bibitem{Becchetti:2019tjy}
Matteo Becchetti, Roberto Bonciani, Valerio Casconi, Andrea Ferroglia, Simone
  Lavacca, and Andreas von Manteuffel.
\newblock {Master Integrals for the two-loop, non-planar QCD corrections to
  top-quark pair production in the quark-annihilation channel}.
\newblock {\em JHEP}, 08:071, 2019.

\bibitem{Chaubey:2019lum}
Ekta Chaubey and Stefan Weinzierl.
\newblock {Two-loop master integrals for the mixed QCD-electroweak corrections
  for $H \rightarrow b\bar{b}$ through a $H t \bar{t}$-coupling}.
\newblock {\em JHEP}, 05:185, 2019.

\bibitem{Becchetti:2017abb}
Matteo Becchetti and Roberto Bonciani.
\newblock {Two-Loop Master Integrals for the Planar QCD Massive Corrections to
  Di-photon and Di-jet Hadro-production}.
\newblock {\em JHEP}, 01:048, 2018.

\bibitem{Bonciani:2016qxi}
Roberto Bonciani, Vittorio Del~Duca, Hjalte Frellesvig, Johannes~M. Henn,
  Francesco Moriello, and Vladimir~A. Smirnov.
\newblock {Two-loop planar master integrals for Higgs$\to 3$ partons with full
  heavy-quark mass dependence}.
\newblock {\em JHEP}, 12:096, 2016.

\bibitem{Bonciani:2016ypc}
Roberto Bonciani, Stefano Di~Vita, Pierpaolo Mastrolia, and Ulrich Schubert.
\newblock {Two-Loop Master Integrals for the mixed EW-QCD virtual corrections
  to Drell-Yan scattering}.
\newblock {\em JHEP}, 09:091, 2016.

\bibitem{Ita:2015tya}
Harald Ita.
\newblock {Two-loop Integrand Decomposition into Master Integrals and Surface
  Terms}.
\newblock {\em Phys. Rev. D}, 94(11):116015, 2016.

\bibitem{Remiddi:2013joa}
Ettore Remiddi and Lorenzo Tancredi.
\newblock {Schouten identities for Feynman graph amplitudes; The Master
  Integrals for the two-loop massive sunrise graph}.
\newblock {\em Nucl. Phys. B}, 880:343--377, 2014.

\bibitem{Gehrmann:2014bfa}
Thomas Gehrmann, Andreas von Manteuffel, Lorenzo Tancredi, and Erich Weihs.
\newblock {The two-loop master integrals for $q\overline{q} \to VV$}.
\newblock {\em JHEP}, 06:032, 2014.

\bibitem{DiVita:2017xlr}
Stefano Di~Vita, Pierpaolo Mastrolia, Amedeo Primo, and Ulrich Schubert.
\newblock {Two-loop master integrals for the leading QCD corrections to the
  Higgs coupling to a $W$ pair and to the triple gauge couplings $ZWW$ and
  $\gamma^*WW$}.
\newblock {\em JHEP}, 04:008, 2017.

\bibitem{vonManteuffel:2020vjv}
Andreas von Manteuffel, Erik Panzer, and Robert~M. Schabinger.
\newblock {Cusp and collinear anomalous dimensions in four-loop QCD from form
  factors}.
\newblock {\em Phys. Rev. Lett.}, 124(16):162001, 2020.

\bibitem{Huber:2019fxe}
Tobias Huber, Andreas von Manteuffel, Erik Panzer, Robert~M. Schabinger, and
  Gang Yang.
\newblock {The four-loop cusp anomalous dimension from the $N=4$ Sudakov form
  factor}.
\newblock {\em Phys. Lett. B}, 807:135543, 2020.

\bibitem{Henn:2019swt}
Johannes~M. Henn, Gregory~P. Korchemsky, and Bernhard Mistlberger.
\newblock {The full four-loop cusp anomalous dimension in $\mathcal{N}=4$ super
  Yang-Mills and QCD}.
\newblock {\em JHEP}, 04:018, 2020.

\bibitem{Bruser:2019auj}
Robin Br\"user, Andrey Grozin, Johannes~M. Henn, and Maximilian Stahlhofen.
\newblock {Matter dependence of the four-loop QCD cusp anomalous dimension:
  from small angles to all angles}.
\newblock {\em JHEP}, 05:186, 2019.

\bibitem{Moch:2018wjh}
S.~Moch, B.~Ruijl, T.~Ueda, J.~A.~M. Vermaseren, and A.~Vogt.
\newblock {On quartic colour factors in splitting functions and the gluon cusp
  anomalous dimension}.
\newblock {\em Phys. Lett. B}, 782:627--632, 2018.

\bibitem{Grozin:2017css}
Andrey Grozin, Johannes Henn, and Maximilian Stahlhofen.
\newblock {On the Casimir scaling violation in the cusp anomalous dimension at
  small angle}.
\newblock {\em JHEP}, 10:052, 2017.

\bibitem{Henn:2016men}
Johannes~M. Henn, Alexander~V. Smirnov, Vladimir~A. Smirnov, and Matthias
  Steinhauser.
\newblock {A planar four-loop form factor and cusp anomalous dimension in QCD}.
\newblock {\em JHEP}, 05:066, 2016.

\bibitem{griffiths1969periods}
Phillip~A Griffiths.
\newblock On the periods of certain rational integrals: Ii.
\newblock {\em Annals of Mathematics}, pages 496--541, 1969.

\bibitem{laporta2000high}
Stefano Laporta.
\newblock High-precision calculation of multiloop feynman integrals by
  difference equations.
\newblock {\em International Journal of Modern Physics A}, 15(32):5087--5159,
  2000.

\bibitem{gelfand1994discriminants}
Israel~M Gelfand, Mikhail~M Kapranov, and Andrei~V Zelevinsky.
\newblock A-discriminants.
\newblock In {\em Discriminants, Resultants, and Multidimensional
  Determinants}, pages 271--296. Springer, 1994.

\bibitem{hibi2017pfaffian}
Takayuki Hibi, Kenta Nishiyama, and Nobuki Takayama.
\newblock Pfaffian systems of a-hypergeometric equations i: Bases of twisted
  cohomology groups.
\newblock {\em Advances in Mathematics}, 306:303--327, 2017.

\bibitem{beilinson2018faisceaux}
Alexander Beilinson, Joseph Bernstein, Pierre Deligne, and Ofer Gabber.
\newblock {\em Faisceaux pervers}.
\newblock Soci{\'e}t{\'e} math{\'e}matique de France, 2018.

\bibitem{Badger:2022hno}
Simon Badger, Matteo Becchetti, Ekta Chaubey, and Robin Marzucca.
\newblock {Two-loop master integrals for a planar topology contributing to pp
  to ttj}.
\newblock 10 2022.

\bibitem{Abreu:2022vei}
Samuel Abreu, Matteo Becchetti, Claude Duhr, and Melih~A. Ozcelik.
\newblock {Two-loop master integrals for pseudo-scalar quarkonium and leptonium
  production and decay}.
\newblock {\em JHEP}, 09:194, 2022.

\bibitem{Duhr:2021fhk}
Claude Duhr, Vladimir~A. Smirnov, and Lorenzo Tancredi.
\newblock {Analytic results for two-loop planar master integrals for Bhabha
  scattering}.
\newblock {\em JHEP}, 09:120, 2021.

\bibitem{Bogner:2007mn}
Christian Bogner and Stefan Weinzierl.
\newblock {Periods and Feynman integrals}.
\newblock {\em J. Math. Phys.}, 50:042302, 2009.

\bibitem{kashiwara2013sheaves}
Masaki Kashiwara and Pierre Schapira.
\newblock {\em Sheaves on Manifolds: With a Short History}, volume 292.
\newblock Springer Science \& Business Media, 2013.

\bibitem{hotta2007d}
Ryoshi Hotta and Toshiyuki Tanisaki.
\newblock {\em D-modules, perverse sheaves, and representation theory}, volume
  236.
\newblock Springer Science \& Business Media, 2007.

\bibitem{arnold2012singularities}
Vladimir~Igorevich Arnold, Aleksandr~Nikolaevich Varchenko, and
  Sabir~Medzhidovich Gusein-Zade.
\newblock {\em Singularities of differentiable maps: Volume II Monodromy and
  asymptotic integrals}, volume~83.
\newblock Springer Science \& Business Media, 2012.

\end{thebibliography}

\end{document}